\documentclass[12pt,preprint]{aastex}

\newcommand{\perval}[2]{{#1\mbox{$^{#2}$}}}
\newcommand{\erg}{\mbox{$\rm\,erg$}\/}

\newcommand{\persec}{\perval{s}{-1}\/}

\newcommand{\cgslum}{\erg~\persec}
\def\gsev {G70.7$+$1.2}
\def\ptsrc{CXO~J200423.4$+$333907} 
\def\lgs {LGS-AO}

\def\kms {\mbox{ km s}^ {-1}}

\def\simlt{\mathrel{\hbox{\rlap{\hbox{\lower4pt\hbox{$\sim$}}}\hbox{$<$}}}}
\def\simgt{\mathrel{\hbox{\rlap{\hbox{\lower4pt\hbox{$\sim$}}}\hbox{$>$}}}}
\def\ra#1#2#3{#1$^{\rm h}$#2$^{\rm m}$#3$^{\rm s}$}
\def\dec#1#2#3{$#1^\circ#2'#3$}
\def\sci#1#2{$#1 \times 10^{#2}$}
\newcommand{\chandra}{{\em Chandra\/}}

\begin{document}
\title{Near-Infrared and X-ray Observations of the Enigmatic \gsev}
\author{P. B. Cameron and S. R. Kulkarni}
\affil{California Institute of Technology, Division Physics,
Math and Astronomy, MC 105-24, Pasadena, CA 91125 \\
Electronic mail: pbc@astro.caltech.edu}
 
\begin{abstract}
We present high resolution imaging of the puzzling radio and optical
nebula \gsev\ with the Keck Observatory's laser guide star adaptive
optics (\lgs) system and the {\em Chandra X-ray Observatory}. The
archival X-ray observations show a hard ($\Gamma \approx 1.8$), low
luminosity ($L_X \approx $ \sci{4}{31} \cgslum) point source at the
center of the nebula. Follow-up \lgs\ near-infrared imaging of the
\chandra\ error circle reveals a relatively bright ($K^{\prime}$
$\approx$ 14 magnitude) counterpart. Both its color and brightness are
consistent with a heavily obscured B-star or possibly a late-G/early-K
giant. The most plausible explanation is that this newly discovered
X-ray source is a non-accreting B-star/pulsar binary powering the
radio and optical nebula. If so, the luminous Be-star discussed in the
literature seemingly embedded in the nebula is not the dominant force
responsible for shaping \gsev. Thus, we suggest that \gsev\ is the
result of two unrelated objects (a B-star X-ray binary and a Be star)
interacting with a dense molecular cloud.  With this explanation we
believe we have solved the mystery of the origin of \gsev.
\end{abstract}

\section{Introduction}
\gsev\ is a compact ($\approx 20$\arcsec) radio and optical nebula in
the Galactic Plane whose origin is controversial
\citep{reich+85,green86,demuizon+88,bally+89}.  The shell-like radio
morphology is accompanied by broad, blue-shifted [O {\sc I}] and [S
{\sc II}] emission lines indicative of an interstellar shock
\citep{demuizon+88,kulkarni92}.  Millimeter CO emission traces this
radio and optical emission, suggesting the shock is interacting with
molecular material \citep{bally+89,phillips93,onello95}.  In addition,
a bright near-infrared (NIR) star appears to be embedded in \gsev, and
it is surrounded by a strong H$\alpha$ reflection of spectral type Be
\citep{becker88,kulkarni92}.

Ironically, it is the plethora of clues that make \gsev\ a perplexing
object, despite its discovery more than two decades ago
\citep{reich+85}.  \gsev\ is unique in that nearly every Galactic
prototype has been proposed to explain it: young supernova remnant,
nova shell, stellar wind bubble, H~{\sc II} region and
Herbig-Haro-like outflow
\citep{reich+85,demuizon+88,green86,becker88}. However, none of these
can explain the low expansion velocities and the non-thermal radio
emission. The only currently proposed consistent theory to explain
these properties is one in which the bright NIR star is paired with an
unseen neutron star to form a Be-radio pulsar binary moving
supersonically through the dense gas \citep{kulkarni92}. In this
model, significant mass loss from the luminous Be-star inflates a
bubble which is filled by a mixture of the stellar wind with energetic
particles and magnetic field from the pulsar.  This combination
creates the non-thermal radio emission coincident with the optical bow
shock of the medium surrounding the system.  This model makes the
prediction that an X-ray source or pulsar should be seen coincident
with the embedded Be-star.

Here, we report on archival X-ray and new Keck \lgs\ observations that
reveal an X-ray source with a NIR counterpart in the center of \gsev\
which is unassociated with the Be-star.  We suggest that the object
known as \gsev\ is the result of the interaction of a luminous Be-star
and an X-ray emitting B-star/pulsar binary with dense molecular
material.  If correct the resulting study of this object will be an
interesting laboratory for the study of plasma processes.  The
observations and results are presented in \S\ref{sec:obs}. In
\S\ref{sec:dis} we discuss the implications of this source as it
relates to resolving the mystery surrounding \gsev.

\section{Observations and Analysis}
\label{sec:obs}
\subsection{X-ray}
\gsev\ was observed 2003 October 11.33 UT with the ACIS-S detector on
{\em Chandra} in the standard, timed exposure mode. The archival data
were analyzed with CIAO version
3.2\footnote{http://www.cxc.harvard.edu/ciao/}. We reprocessed the
level 1 events from the {\em Chandra} X-ray Center (CXC) in order to
make use of the latest calibration and removed pixel randomization.
The level 2 event file was created by filtering grades 0,2,3,4,6 and
good-time intervals.  The total exposure time after filtering periods
higher than 3-$\sigma$ above the mean background level was 37.6\,ksec.

Diffuse emission and a point source (hereafter \ptsrc;
Figure~\ref{fig:combo}) are detected at the position of \gsev\ (as
first noted by \citealt{arz04}). We compared the positions of 27 X-ray
sources on the S3 chip with counterparts in the 2MASS point source
catalog to correct the native astrometry \citep{cutri03}. This
comparison showed evidence for a small systematic shift,
$\Delta\alpha_{\rm 2MASS-CXO} =$ -0\farcs13 $\pm$ 0\farcs11,
$\Delta\delta_{\rm 2MASS-CXO} =$ -0\farcs08 $\pm$ 0\farcs11. The
best-fit position of \ptsrc\ including this offset is
$\alpha$(J2000)$=$\ra{20}{04}{23.430} and
$\delta$(J2000)$=$\dec{33}{39}{06\farcs73} with measurement uncertainty of
0\farcs03 and 0\farcs08 in each coordinate, respectively. Combining
the measurement and transformation errors in quadrature gives an
uncertainty of 0\farcs18 (1-$\sigma$) for the X-ray position of
\ptsrc\ relative to 2MASS. This position lies 3\farcs6 from the nearby
luminous NIR star, which implies the two are not associated
\citep{kulkarni92,arz04}.

We applied the adaptive smoothing algorithm CSMOOTH to highlight
the diffuse emission after subtraction of \ptsrc\ and produced a
flux-calibrated image by applying an exposure
map as outlined in the CIAO threads. The contours of this emission are
overlaid on an NIR image (see \S\ref{sec:lgsao}) of the nebula in
Figure~\ref{fig:combo}. Evidently, most of the diffuse X-ray emission
is not coincident with the diffuse NIR emission.

We extracted photons within a 1\farcs5 circle (corresponding to 90\%
of the expected counts at 1.4\,keV) around \ptsrc\ to perform spectral
and variability analyses.  The source contains only 33$^{+7}_{-6}$
counts.  Upon examination of regions both inside the diffuse emission
and in a source-free area, we expect only two of these to be
background photons.  A Kolmogorov-Smirnov test shows that the arrival
times of the source photons differed from a constant rate at only the
$\approx$ 1-$\sigma$ level, thus the source cannot be considered
variable.

We begin our spectral analysis by noting that all the photons from
\ptsrc\ fall in the range 1.0--4.5\,keV, suggesting a hard
spectrum. After calculating the response matrix and effective area of
this portion of the CCD, we fit an absorbed power-law model to the
spectrum using Cash statistics (due to the limited number of counts;
\citealt{cash79}). The best-fit parameters in Table~\ref{tab:fit} show
a relatively hard photon index, $\Gamma \approx 1.8$, and low
luminosity, $L_X \approx 4 \times 10^{31}$\,erg\,s$^{-1}$
(2.0--10.0\,keV) for an assumed distance of 4.5\,kpc
\citep{bally+89}. These values are consistent with known accreting
neutron stars in quiescence (e.g. \citealt{rutledge01,campana05}).
In addition, we fit two absorbed power-law models with fixed
parameters (see Table~\ref{tab:fit}). The first has the photon index
set to a typical value for quiescent neutron stars, $\Gamma=2$, while
the second has the column density fixed to the best fit value of the
diffuse emission, $N_H$ = $1.0 \times 10^{22}$\,cm$^{-2}$ (see below).

The probability of finding a source as bright or brighter than \ptsrc\
within the extent of \gsev\ can be determined from the local source
density. A WAVDET analysis of the active CCDs (ACIS-I2,3 and
ACIS-S1,2,3,4) finds that 12 sources are as bright or brighter than
\ptsrc. The inferred density is then $\approx$ 112
sources/deg$^{2}$. This density is consistent with observations taken
as part of the ChaMPlane Survey \citep{grindlay05}, which predicts
$\approx$ 100 sources/deg$^{2}$ with fluxes as bright or brighter than
\ptsrc\ \citep{hong05}.  Consequently, there is a 0.3\% probability
that such a source would be found within \gsev\ by chance.

The diffuse emission presented enough counts for basic spectroscopy
with $\chi^2$ statistics.  We extracted events from a region of
dimension $\approx30$\arcsec$\times 30$\arcsec\ surrounding the
diffuse emission (excluding the point source) and a source-free
background region immediately east of the nebula with the same shape.
This yielded $690 \pm 26$ source counts, of which $\approx$ 320 are
expected to be due to the background.  The resulting source plus
background photons were grouped such that each bin contained at least
25 counts.  

The background subtracted spectrum was analyzed using
XSPECv11\footnote{See http://heasarc.nasa.gov/docs/xanadu/xspec/}.  We
fit two models modified by absorption to the spectrum: a power-law and
a Raymond-Smith plasma (see Table~\ref{tab:fit}).  The unphysically
steep photon index of the power-law model and the lower $\chi^2_\nu$
value lead us to adopt the Raymond-Smith model for the remainder of
our analysis.  The derived value of $N_H$ is reasonably consistent
with that of \ptsrc\ and estimated value of
$1.25\times10^{22}$\,cm$^{-2}$ from \citet{dickey90}.  Integrating
this model over the 0.5--2.5\,keV bandpass implies a luminosity of 8.1
$\times$ 10$^{32}$\cgslum\ at the distance of \gsev, although it is
not clear that this emission is associated with the nebula (see
\S\ref{sec:dis}).

\subsection{Near-Infrared LGS-AO}
\label{sec:lgsao}
\gsev\ was observed under photometric conditions on 2005 April 30 UT
with Laser Guide Star Adaptive Optics (\lgs; \citealt{lgs1,lgs2}) on
the Keck II telescope and the Near-Infrared Camera 2 (NIRC2). We
imaged the field in the $J$, $H$ and $K^{\prime}$-bands with the wide
camera of NIRC2, which provides a $\approx$40\arcsec$\times$40\arcsec\
field of view and a $\approx$ 0\farcs04 pixel scale.  The $H$ and
$K^{\prime}$-band data sets consisted of five frames in each
band. Each frame was exposed for 5\,sec with 10 additions performed on
the chip at five dither positions separated by $\approx$
30\arcsec. The $J$-band data consisted of two images at the center of
the chip.

Each frame was flat-fielded, background subtracted, and repaired for
bad pixels using custom PyRAF software\footnote{PyRAF is a product of
Space Telescope Science Institute, which is operated by AURA for
NASA.}. We then performed a second round of sky subtraction using a
median combination of similarly processed frames of a nearby field. We
used these processed images of \gsev\ for photometric analysis, but
produced a separate set of images for astrometry due to optical
distortion in the NIRC2 camera. The distortion in the second set was
corrected using algorithms derived from the preshipment review
documents\footnote{available at
http://www2.keck.hawaii.edu/inst/nirc2/} with the IDL procedure
provided by the Keck Observatory\footnote{See
http://www2.keck.hawaii.edu/optics/lgsao/software/}. The correction
does not conserve flux, and thus is not suitable for photometry.

We registered a median combination of the distortion corrected
$H$-band frames to the 2MASS point source catalog using 8 stars that
were not over-exposed.  We find residuals of 0\farcs04 and 0\farcs09
in right ascension and declination, respectively.  Registering the $J$
and $K^{\prime}$-band frames to this image yielded negligible
residuals. Combining these errors with those in the X-ray position of
\ptsrc\ yields an uncertainty of 0\farcs19 (1-$\sigma$) of the X-ray
image with respect to the NIR images.  Figure~\ref{fig:combo} shows
the registered $H$-band frame with the Chandra error circle (99\%
confidence). We clearly identify a single bright NIR counterpart in
all filters within the X-ray error circle.  The best fit position of
this source is $\alpha$(J2000)$=$\ra{20}{04}{23.446} and
$\delta$(J2000)$=$\dec{33}{39}{06\farcs62} with an uncertainty of
0\farcs04 and 0\farcs09 (relative to 2MASS), respectively.  The
centroiding errors are negligible. This position lies 0\farcs23 from
the {\em Chandra} position.

We performed aperture photometry of the source in each band on each
individual frame relative to 2MASS stars in the field with the IRAF
package APPHOT. We assume that the color term used to transform from
the 2MASS $K_s$ filter to the $K^{\prime}$ is negligible for our
purposes. The uncertainties were determined with the 2MASS photometric
uncertainty, the standard deviation of the zero-point determinations
from the same 2MASS star in multiple frames and the photometric error
of the NIR source itself added in quadrature. We find magnitudes of $J
= 15.56 \pm 0.09$, $H = 14.51 \pm 0.11$ and $K^{\prime} = 13.97 \pm
0.06$.

The probability of finding a star with $K^{\prime} \approx 14$ magnitude in
our \chandra\ error circle by chance is very low. To quantify this we
extracted all sources present in the 2MASS catalog within 20\arcmin\
of \gsev. We find that the differential number of sources per
magnitude per square arcsecond is well described by a single power-law
with index 0.35 over the magnitude range 3 $< K_{\rm s} < $ 15. We can
conservatively assume (based on Galactic star count models by
\citealt{nakajima00}) that this can be extrapolated to our 5-$\sigma$
detection limit of $m_{K^{\prime}} \approx 20.0$ magnitude.  From this
we calculate that there is a $\approx$ 25\% percent chance of finding
a source brighter than our detection limit in a circular region with a
0\farcs49 (99\% confidence) radius. However, the probability of
finding a source with $K_{s} = 14.0$ magnitude or brighter is $\simlt$
0.1\%. Thus it is unlikely that our NIR counterpart is drawn from the
background population, and we assume that it associated with \ptsrc.

\begin{figure}
\plotone{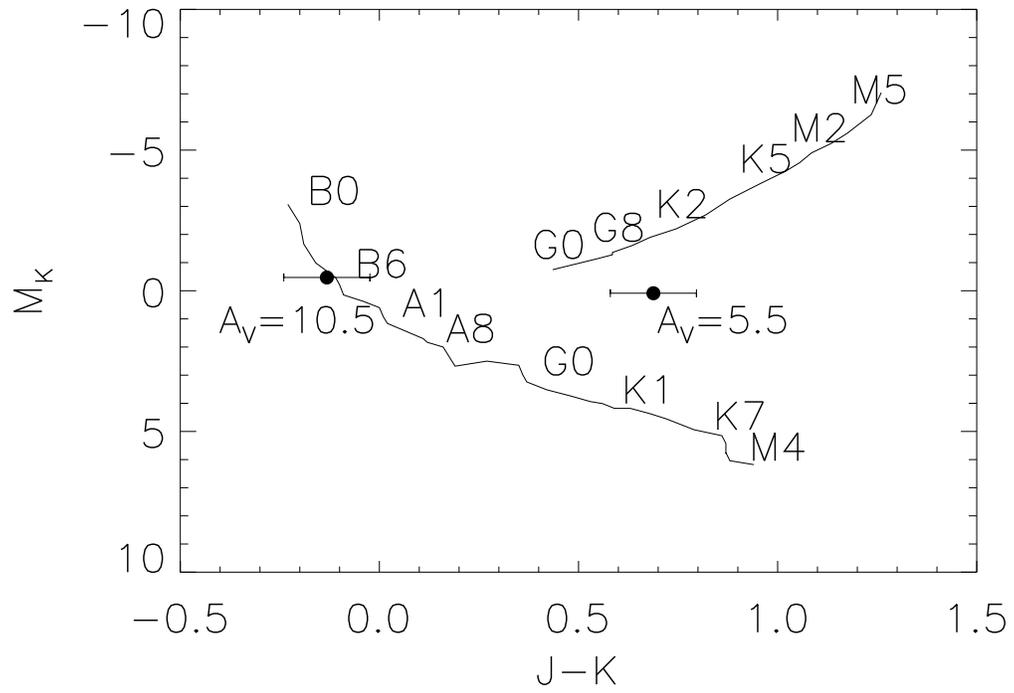}
\caption{Color magnitude diagram using data from \citet{bessell88}.
  Filled circles show the IR counterpart for $A_V = 5.5$ and $A_V =
  10.5$ at a distance of 4.5\,kpc .}
\label{fig:cmd}
\end{figure}

The key issue in determining the nature of this source is the assumed
extinction. Based on the colors of the luminous NIR star,
\citet{becker88} estimate $A_V \approx 5.6$. This agrees well with the
value of $A_V \approx 5.4$ obtained by taking $N_H$ as determined from
the spectrum of the diffuse X-ray emission and translating it into
extinction \citep{predehl95}. If we plot the NIR counterpart on a
color-magnitude diagram (see Figure~\ref{fig:cmd}) using this
reddening we find that the star is consistent with a late G/early K
giant spectral type at a distance of $\approx 11$\,kpc. Consequently,
the star is under luminous if it is associated with \gsev\ at a
distance of 4.5\,kpc.

The spectral fitting of the point source spectrum itself, albeit with
poor statistics, implies a higher extinction of $A_V =
8.4^{+6.4}_{-5.5}$. This allows for the possibility that the NIR
counterpart is a heavily obscured main sequence B-star with $A_V
\approx 10.0$ at the distance of \gsev.  We prefer this interpretation
when we consider the probabilistic arguments and existing
multifrequency observations of \gsev\ (see \S\ref{sec:dis}).

\section{Discussion and Conclusions}
\label{sec:dis}
We identify a low-luminosity, hard X-ray point source with a NIR
counterpart at the center of \gsev\ using high resolution imaging.
Both the measured X-ray luminosity, $L_X \approx 4 \times
10^{31}$\,\cgslum, and the photon index, $\Gamma \approx 1.8$, of
\ptsrc\ are consistent with quiescent neutron star systems
(e.g. \citealt{rutledge01,campana05}). The magnitude and
$J$-$K^{\prime}$ color in combination with the X-ray column density
suggests the NIR counterpart is either an evolved background star or a
heavily extincted B-star. However, an isolated background
late-G/early-K giant cannot explain the observed X-ray flux. These
stars have deep convective zones that power coronal X-ray emission,
but it is typically $\simlt 10^{31}$\cgslum \citep{gudel04}.  This is
an order of magnitude below the required $L_X \approx $
\sci{2.5}{32}\cgslum calculated assuming the observed X-ray flux at a
distance of 11\,kpc. In addition, spectral types later than B2 have
have observed X-ray luminosities $\simlt 10^{31}$\cgslum
\citep{berghofer97}.  This suggests that the NIR source and \ptsrc\
constitute an X-ray binary, and probabilistic arguments suggest that
this binary is associated with \gsev.

\begin{figure}
\plotone{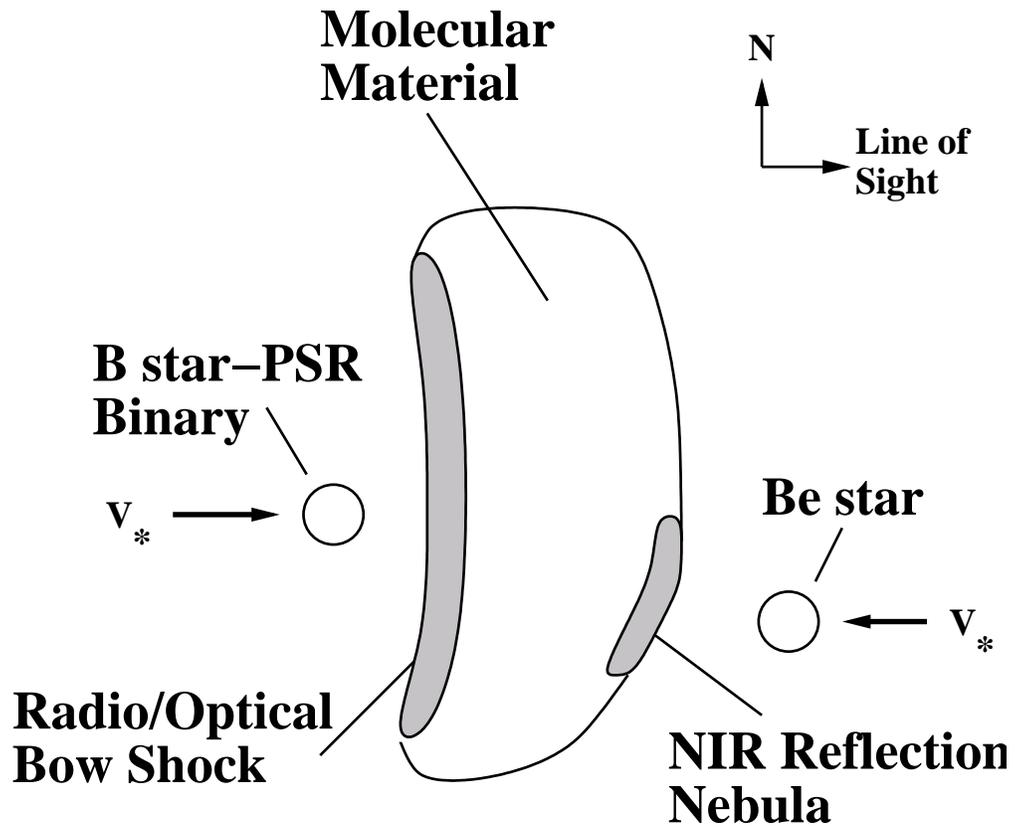}
\caption{Diagram of the geometry of \gsev. The newly discovered X-ray
binary moves into the far side of the molecular material powering the
radio/NIR/optical nebula, whereas the Be-star creates a reflection
nebula on the near-side. See the text for details.}
\label{fig:diagram}
\end{figure}

A simple geometric model can explain the existing multifrequency data
(Figure~\ref{fig:diagram}).  The velocity of the molecular gas as
measured by CO observations is $5\kms$ with respect to the local
standard of rest \citep{bally+89}. The stellar H$\alpha$ line profile
from the bright Be-star is redshifted with respect to the CO with a
velocity of 20--60$\kms$, while H$\alpha$ reflected by dust in the
eastern region is also redshifted with respect to the CO, but is
blueward of the stellar H$\alpha$ by 10--50$\kms$. This implies that
the bright NIR star is moving into the cloud, away from the
Earth. However, the [O I] and H$\alpha$ throughout the rest of nebula
traces the non-thermal radio emission and is uniformly blue-shifted by
10--120$\kms$ with respect to CO, suggesting that the source
responsible for the shock is moving into the cloud, toward the Earth.

The cloud size, as inferred from CO, is 3$D_{4.5}$\,pc on the sky,
where $D_{4.5}$ is the distance to \gsev\ in units of 4.5\,kpc.  If
the cloud is roughly spherical and has $n_H \sim 10^3$\,cm$^{-3}$,
then objects will have an additional $\approx$ 5 magnitudes of
extinction with respect to objects on the near side. Thus, a natural
explanation for the geometry of \gsev\ is that the bright Be-star is
moving into the near side of the cloud creating a reflection nebula,
while on the far side, a heavily extincted B-star/pulsar binary is
moving into the cloud creating a bow shock and powering the nebula
(Figure~\ref{fig:diagram}).

\begin{figure}
\plotone{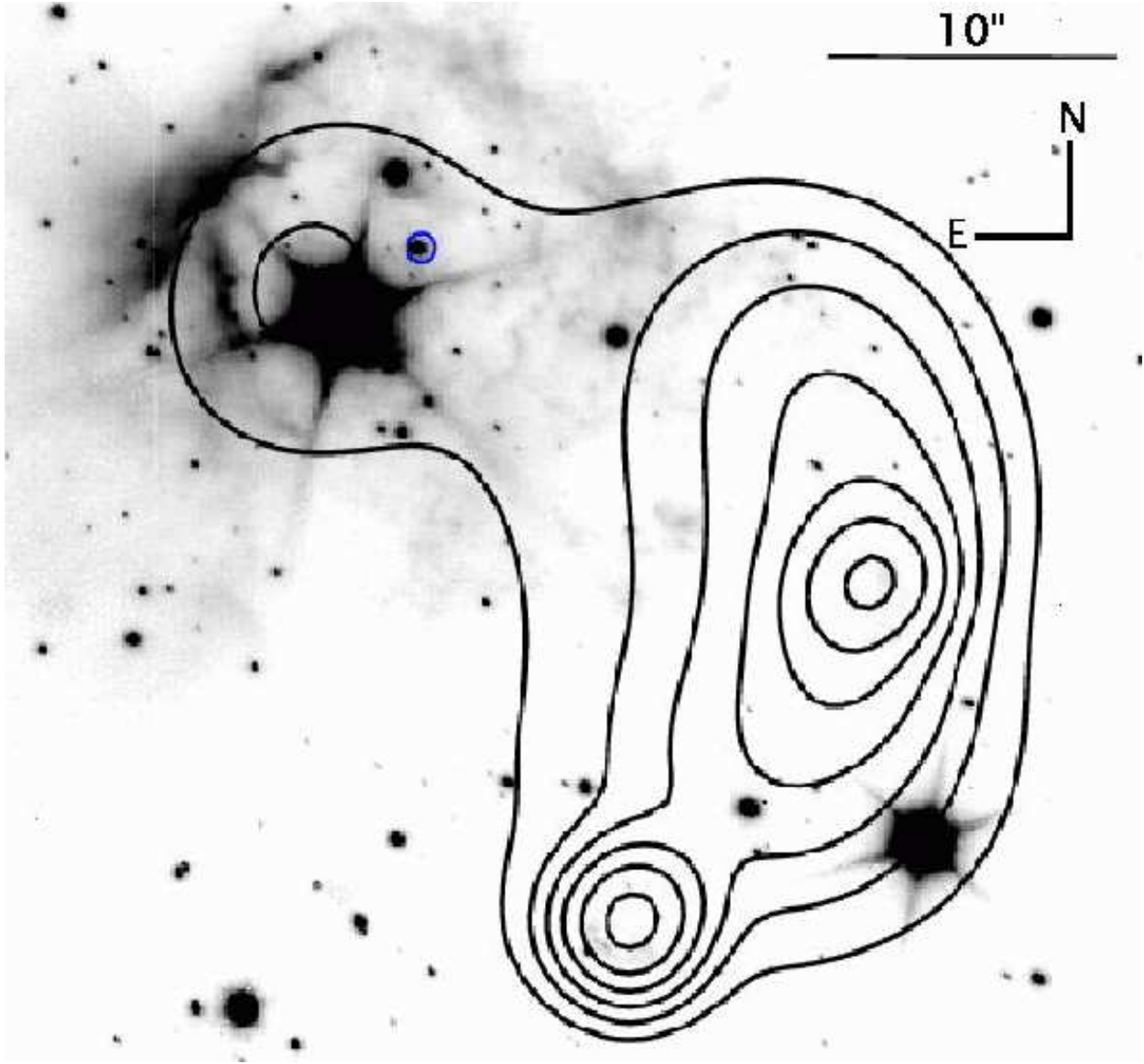}
\epsscale{0.8}
\caption{H-band image of \gsev\ with contours of the adaptively
smoothed X-ray emission (black lines) and the \chandra\ 99\%
confidence (0\farcs49) error circle (blue circle). The X-ray
contours are logarithmically spaced between 10\% and 90\% of the peak
emission.}
\label{fig:combo}
\end{figure}

One remaining puzzle is the origin and impact of the hot gas powering
the diffuse X-ray emission.  Figure~\ref{fig:combo} shows that the
radio/optical and diffuse X-ray morphologies are substantially
different, and the peak of the diffuse X-ray emission is separated
$\approx 20$\arcsec from the center of the radio/optical emission
(which contains the Be-star and X-ray binary). Thus, it is apparent
that this hot gas does not play an important dynamic nor, given the
its luminosity is $\sim 10^{32}$\cgslum, energetic role in shaping
\gsev. Two viable explanations for the origin of the hot gas are,
given the quasi-shell like morphology, that it is the result of a
previous explosive event that the X-ray binary is overtaking or it may
be unassociated with the \gsev. In any case, the origin of this plasma
--- either related or unrelated to \gsev\ --- is unknown.

The definitive proof of the proposed model (Figure~\ref{fig:diagram})
would be the discovery of a pulsar associated with \gsev. A search for
pulsations with the Green Bank Telescope at 2.2\,GHz is underway. If a
pulsar is found, \gsev\ will be an important laboratory for studying
plasma processes taking place between the pulsar/B-star wind and the
interaction of that mixture with the cold molecular gas.

\acknowledgements We thank A. Kraus for useful discussions.  This work
is supported in part by grants from the National Science Foundation
and NASA. The W. M. Keck Observatory is operated as a scientific
partnership among the California Institute of Technology, the
University of California, and the National Aeronautics and Space
Administration. The Observatory was made possible by the generous
financial support of the W. M. Keck Foundation. The authors wish to
recognize and acknowledge the very significant cultural role and
reverence that the summit of Mauna Kea has always had within the
indigenous Hawaiian community. We are most fortunate to have the
opportunity to conduct observations from this mountain.

\bibliographystyle{apj}

\clearpage

\begin{deluxetable}{lccccc}
\tablecaption{\label{tab:fit} X-ray Spectral Fits.}
\tablehead{
\colhead{Model} & \colhead{$N_H$} & \colhead{$\Gamma$/$k_BT$}  & 
\colhead{Flux} & \colhead{$\chi^2/\nu$} \\ 
 & \colhead{($10^{22}$~cm$^{-2}$)} &  \colhead{(keV)} & 
\colhead{($10^{-14}$~erg~cm$^{-2}$~s$^{-1}$)} & \\
\colhead{(1)}& \colhead{(2)}& \colhead{(3)}& \colhead{(4)}& \colhead{(5)}\\
}
\startdata
Diffuse Emission \\
\hline
Power-law &  $0.70^{+0.19}_{-0.13}$ & $4.4^{+0.9}_{-0.5}$   & $31^{+22}_{-11}$ & 18.2/13\\
Raymond-Smith Plasma &  $1.04^{+0.08}_{-0.07}$ & $0.71^{+0.05}_{-0.07}$   & $33^{+9}_{-5}$ & 9.75/13\\
\hline
\ptsrc \\
\hline
Power-law & $1.5^{+1.1}_{-1.0}$ & 1.8$^{+1.2}_{-1.1}$ & 1.7$^{+5.6}_{-1.7}$ & ---\\
Power-law ($\Gamma=2.0$)& $1.6^{+0.5}_{-0.4}$ & (2.0) & 1.6$^{+0.5}_{-0.4}$ & ---\\
Power-law ($N_H=1\times10^{22}$)& (1.0) & $1.3^{+0.5}_{-0.4}$ & $2.0^{+1.0}_{-0.7}$ & ---\\
\enddata
\tablecomments{ All errors are 68\% confidence levels. Values in parentheses are held fixed.
(1)~--~Absorbed spectral model.
(2)~--~Best-fit column density.
(3)~--~Measured photon index for power-law models and $k_BT$ for the 
       Raymond-Smith plasma.
(4)~--~The unabsorbed flux in the 0.5--2.5\,keV band for the diffuse emission 
       and 2--10\,keV band for \ptsrc.
(5)~--~The value of $\chi^2$ for diffuse emission models and the number 
       of degrees of freedom, $\nu$. This column is not applicable to
       \ptsrc\ since the spectral fitting was performed with Cash 
       statistics.
}
\end{deluxetable}

\clearpage

\clearpage

\clearpage

\end{document}